\documentclass[a4paper,11pt]{article}
\usepackage{pos}

\def\beq{\begin{equation}}
\def\eeq{\end{equation}}
\def\bea{\begin{eqnarray}}
\def\eea{\end{eqnarray}}

\def\ii{\imath 0}
\def\qon#1{q_{#1,0}^{(+)}}

\def\qb{\mathbf{q}}

\def\ket#1{|{#1}\rangle}

\usepackage{color}
\usepackage{ulem}

\title{Quantum Algorithm for Querying Causality of Multiloop Scattering Amplitudes}
\ShortTitle{Quantum Algorithm for Querying Causality}

\author*[a]{Selomit Ram\'irez-Uribe}

\affiliation[a]{Instituto de F\'isica Corpuscular, Universitat de Val\`encia -- Consejo Superior de Investigaciones Cient\'ificas,
  Parc Cient\'ific, Paterna, Spain}

\affiliation[a]{Facultad de Ciencias de la Tierra y el Espacio, Universidad Aut\'onoma de Sinaloa,\\
Ciudad Universitaria, Culiac\'an, M\'exico}

\affiliation[a]{Facultad de Ciencias F\'isico-Matem\'aticas, Universidad Aut\'onoma de Sinaloa,\\
Ciudad Universitaria, Culiac\'an, M\'exico}

\emailAdd{norma.selomit.ramirez@ific.uv.es}

\abstract{
The first application of a quantum algorithm to Feynman loop integrals is reviewed. 
The connection between quantum computing and perturbative quantum field theory is feasible due to fact that the two on-shell states of a Feynman propagator are naturally encoded in a qubit. 
The particular problem to be addressed is the identification of the causal singular configurations of multiloop Feynman diagrams. 
The identification of such configurations is carried out through the implementation of a modified Grover's quantum algorithm for querying multiple solutions over unstructured datasets.
}

\FullConference{%
  41st International Conference on High Energy physics - ICHEP2022\\
  6-13 July, 2022\\
  Bologna, Italy
}

\begin{document}
\maketitle
\section{Introduction}
The main challenge in perturbative Quantum Field Theory at colliders is the computation of multiloop scattering amplitudes. In order to face this difficult task, a novel methodology has been developed, the loop-tree duality (LTD)~\cite{Catani:2008xa,Rodrigo:2008fp,Bierenbaum:2010cy,Bierenbaum:2012th,Buchta:2014dfa,Tomboulis:2017rvd,Runkel:2019yrs,Capatti:2019ypt} which opens any Feynman loop diagram into a sum of connected trees. The effort to deepen in the LTD is reflected in their appealing features~\cite{Buchta:2015wna,Hernandez-Pinto:2015ysa,Sborlini:2016gbr,Sborlini:2016hat,Jurado:2017xut,Driencourt-Mangin:2017gop,Driencourt-Mangin:2019yhu,Plenter:2020lop,Aguilera-Verdugo:2019kbz,Prisco:2020kyb}, furthermore, the latest progress in this framework includes a clever reformulation which was first presented in Ref.~\cite{Verdugo:2020kzh}. This result allowed to exploit its most remarkable property, the existence of a manifestly causal representation, opening the path for a significant evolution~\cite{snowmass2020,Aguilera-Verdugo:2020kzc,Aguilera-Verdugo:2020nrp,Ramirez-Uribe:2020hes,Sborlini:2021owe,Bobadilla:2021pvr,TorresBobadilla:2021ivx,Aguilera-Verdugo:2021nrn,Ramirez-Uribe:2021ubp,RamirezUribe:2021nlj,Uribe:2021cdb,Ramirez-Uribe:2022gxz}. 

The possibility to work in a causal LTD scenario, in which noncausal singularities are absent, gives the advantage to work with more numerically stable integrands~\cite{Aguilera-Verdugo:2020kzc,Ramirez-Uribe:2020hes}.
To achieve a causal LTD representation it is required to identify the causal configurations of the multiloop topology of interest.
In this presentation, we review the solution presented in Ref.\cite{Ramirez-Uribe:2021ubp,RamirezUribe:2021nlj,Uribe:2021cdb} for the unfolding of those configurations fulfilling causal conditions by the application of a modified Grover's quantum algorithm~\cite{Grover:1997fa} for querying multiple solutions over unstructured databases~\cite{Boyer:1996zf}. A variational quantum eigensolver approach has recently been presented in Ref.~\cite{Clemente:2022nll}.

\section{LTD framework}
The LTD representation of any multiloop scattering amplitude is computed by the iterative evaluation of the Cauchy's residue theorem in the terms of nested residues~\cite{Verdugo:2020kzh,Ramirez-Uribe:2020hes}.	
Regarding the causal LTD expression, it is obtained by summing over all the nested residues and followed by an ingenious rearrangement~\citep{Aguilera-Verdugo:2020kzc,Ramirez-Uribe:2020hes} that allows to achieve the following causal dual form,
\begin{equation}
{\cal A}_D^{(L)} = \int_{\vec \ell_1 \ldots \vec \ell_L} 
\frac{1}{x_n} \sum_{\sigma  \in \Sigma} \frac{{\cal N}_{\sigma(i_1, \ldots, i_{n-L})}}{\lambda_{\sigma(i_1)}^{h_{\sigma(i_1)}} \cdots \lambda_{\sigma(i_{n-L})}^{h_{\sigma(i_{n-L})}}}
+ (\lambda_p^+ \leftrightarrow \lambda_p^-)~,
\label{eq:AD}
\end{equation}
where $x_n = \prod_n 2\qon{i}$.
This expression is written in terms of causal propagators, $1/\lambda_p^\pm$, with $\lambda_p^\pm = \sum_{i\in p} \qon{i} \pm k_{p,0}$ , on-shell energies $\qon{i}=\sqrt{\qb_i^2+m_i^2-\ii}$ and $k_{p,0}$ a linear combination of the energy components of external momenta.
After every propagator in the set $p$ is set on shell, and according the sign of $k_{p,0}$, either $\lambda_p^-$ or $\lambda_p^+$ becomes singular.
The collection of entangled causal propagators is encoded $\Sigma$, representing all combinations of compatible causal thresholds.

To work in a general context, we recall the concepts of edges and eloops~\cite{TorresBobadilla:2021ivx,Sborlini:2021owe}. Edges are defined as the union of an arbitrary number of propagators connecting two interaction vertices.
The algorithm and the implementation is presented in terms of eloops, or loops made of edges, given that in a causality scenario the causal singular configurations only consider the setting of having all the momentum flow of all the propagators in an edge aligned in the same direction.

\section{ Feynman loop integrals in a quantum computer} 

The feasibility of associating quantum computing and Feynman loop integrals is due to the fact that the two on-shell states of a Feynman propagator can be stored in a qubit. The initial momentum flow for a specific topology is represented with the state $\ket{1}$, whereas $\ket{0}$ encodes those states with inverse flow orientation. 

\subsection{Modified Grover's quantum algorithm}
\begin{figure}[t]
\raisebox{15pt}{\includegraphics[scale=0.2]{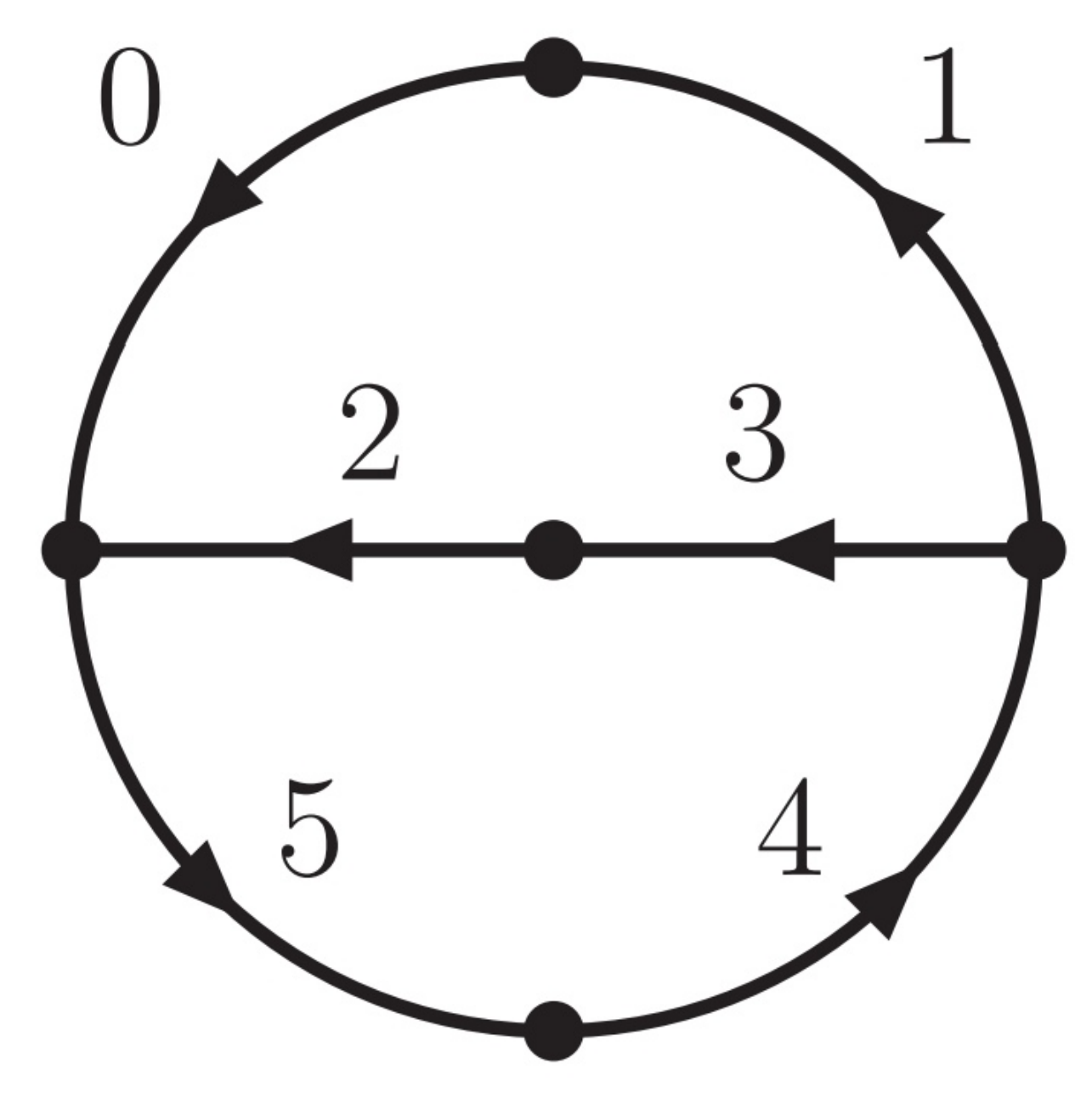}} \qquad
\includegraphics[scale=0.30]{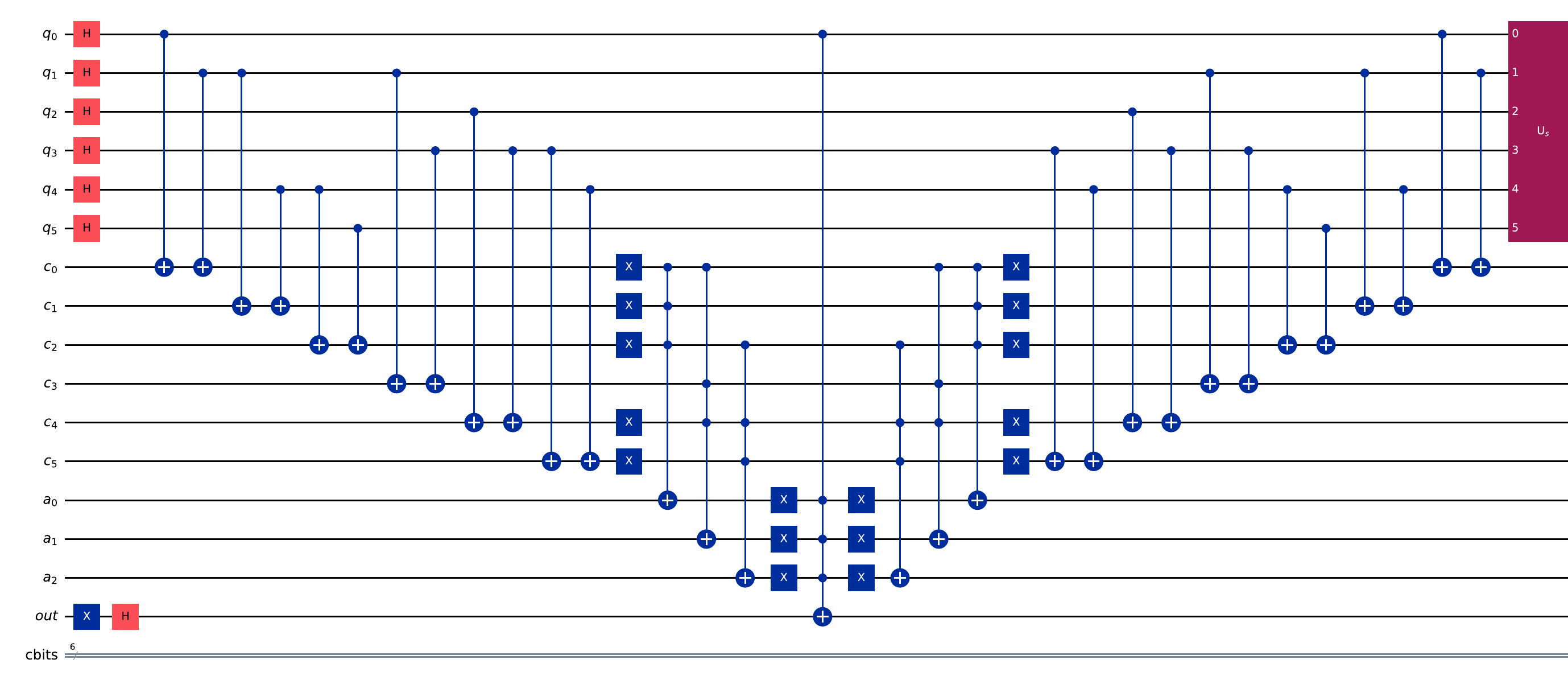}
\centering
\caption{Six-edge two-eloop topology (left) and its corresponding quantum circuit (right) used to bootstrap its causal configurations. 
\label{fig:two-eloop}}
\end{figure}

The global structure of Grover's quantum algorithm relies in three main components: the uniform superposition of all possible states, an oracle and a diffuser operator. Regarding Feynman loop integrals, the states refer to the associated configuration of the diagram. The total number of possible configurations is $N=2^n$, with $n$ the number of edges. Additionally, the causal configurations represent the winning states and the noncausal configurations the orthogonal states.

The uniform superposition of all the $N$ states is written in terms of the uniform superposition of the causal $\left(~\ket{w}~\right)$ and noncausal $\left(~\ket{q_\perp}~\right)$ states as: $\ket{q} = \cos \theta \, \ket{q_\perp} + \sin\theta \, \ket{w}$. The mixing angle between the causal and noncausal states is denoted by $\theta = \arcsin \sqrt{r/N}$, with $r$ the number of causal states. 
The oracle operator, $U_w$, flips the state $\ket{x}$ if $x\in w$, and does nothing otherwise; the diffuser operator, $U_q$, amplifies the probability of these elements by performing a reflection around the initial state $\ket{q}$. 
The iterative application of both operators $t$ times gives $(U_q U_w )^t \ket{q} = \cos \theta_t \, \ket{q_\perp } +  \sin \theta_t \, \ket{w}$, with $\theta_t = (2t +1) \, \theta$.

To define a proper number of iterations, $\theta_t$ has to fulfill that the noncausal state probabilities are much smaller than causal state probabilities. 
Based on this condition, if the initial mixing angle is less or similar to $\pi /6$, Grover's quantum algorithm is considered a feasible framework. 

Regarding the identification of causal singular configurations of multiloop Feynman diagrams, it is known from classical~\cite{Aguilera-Verdugo:2020kzc,Ramirez-Uribe:2020hes} and quantum~\cite{Ramirez-Uribe:2021ubp} computations that in most cases they do not fit in a favorable scenario for a direct application of Grover's algorithm.
A suitable adjustment is given by fixing one qubit to reduce the number of causal solutions, taking advantage that given one causal solution the mirror configuration, is also a causal solution~\cite{Ramirez-Uribe:2021ubp,RamirezUribe:2021nlj,Uribe:2021cdb,Ramirez-Uribe:2022gxz}; if the case requires it, also the total number of states can be increased by introducing an ancillary qubit in the $\ket{q}$ register~\cite{Nielsen2000}. 

The modified Grover's quantum algorithm requires three registers and one extra qubit used as a marker in the oracle. The $n$ edges are encoded in $n$ qubits $q_i$; 
the binary clauses compare the orientation of two adjacent edges through $c_{ij} \equiv (q_i = q_j)$ and $\bar c_{ij} \equiv (q_i \ne q_j)$ with $i,j \in\{0, \ldots, n-1\}$, which are stored in $\ket{c}$.
The loop clauses encoded in the register $\ket{a}$ combine the qubits from $\ket{c}$ to test if any subloop configurations generate a cyclic circuit.

The structure of the algorithm takes as a first step the initialization of all the registers involved. 
The qubits encoding the edges are set in a uniform superposition, $\ket{q} = H^{\otimes n} \ket{0}$; the registers $\ket{a}$, $\ket{c}$ are set to the state $\ket{0}$; and the Grover's marker is set to, $\ket{out_0} = \left(\ket{0} - \ket{1}\right)/\sqrt{2}$.

The implementation of $\bar c_{ij}$ needs two CNOT gates acting between $q_i$, $q_j$ and a qubit in $\ket{c}$. In the case of $c_{ij}$, an extra XNOT gate is used to operate on the respective qubit in $\ket{c}$. The loop clauses are set by combining binary clauses and implemented through multicontrolled Toffoli gates.

To test the causal conditions we define $f(a,q)$. If the causal conditions are satisfied then $f(a,q) = 1$, if not $f(a,q) = 0$. 
The implementation of the oracle, to identify and to mark, is given by
$U_w \ket{q} \ket{c} \ket{a} \ket{out_0} = \ket{q} \ket{c} \ket{a} \ket{out_0 \otimes f(a,q)}$, where $\ket{out_0 \otimes f(a,q)}$ is $-\ket{out_0}$ if $q \in w$, and $\ket{out_0}$ if $q \not\in w$. 
The process continues by applying the oracle operations in the inverse order.
As a final step before measuring, the amplification of the probabilities is applied by implementing $U_q$ to $\ket{q}$. The function of the diffuser operator is taken from IBM Qiskit (\texttt{https://qiskit.org/}). 

\subsection{Two-eloop topology with six edges}

 \begin{figure}[t]
\includegraphics[width=\textwidth]{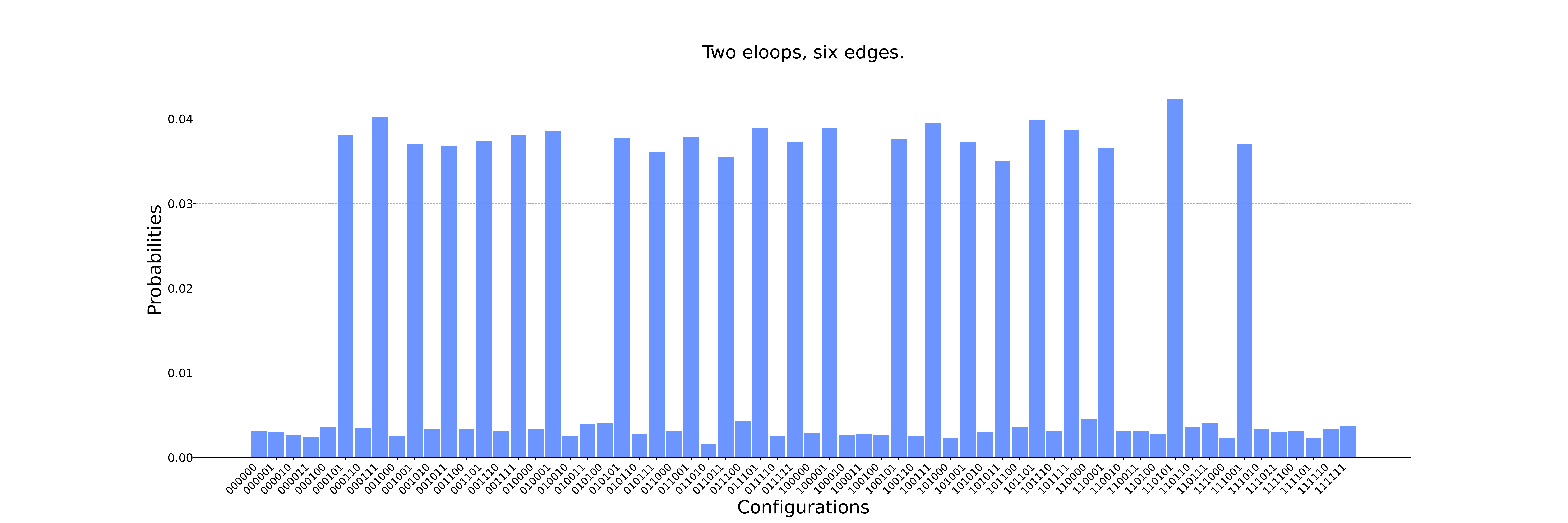}
\centering
\caption{Probabilities of causal and noncausal configurations of a two-eloop topology with six edges. 
\label{fig:output}}
\end{figure}
The implementation of the algorithm described in the previous section is illustrated with the two-eloop topology depicted in Fig.~\ref{fig:two-eloop} (left), consisting of six edges.
From a classical computation it is found that $\theta \thickapprox \pi/3$, therefore, a suitable modification is required. 
Halving the number of solution leads to $\theta \thickapprox \pi/5$, which allows to implement Grover's quantum algorithm, furthermore, under this arrangement only one iteration is needed to achieve a desire probability amplification.

The total number of qubits required for the implementation of the algorithm is sixteen, six for encoding the edges,  six to store the binary clauses, three to test the causal conditions and the one standing as the Grover's oracle marker.
The six-edge two-eloop diagram (Fig.~\ref{fig:two-eloop}) requires to test three subloops:
$a_0 = \neg \left( c_{01} \wedge c_{14}\wedge c_{45}\right)~, \quad a_1 = \neg \left( c_{01} \wedge \bar c_{13} \wedge c_{23} \right)~$ and $~a_2 = \neg \left( c_{23} \wedge c_{34} \wedge c_{45} \right)$. 

The Boolean condition marking the causal configurations is $f^{(2)}(q,a)=(a_0\wedge a_1 \wedge a_2)\wedge q_0$, where $q_0$ is fixed.
The quantum circuit associated to the algorithm is depicted in Fig.~\ref{fig:two-eloop} (right), successfully implemented in the IBM Qiskit quantum simulator {\tt{qasm}}. The output shown in Fig.~\ref{fig:output} successfully identifies the expected twenty three causal configurations, corresponding to the forty six causal states when including the mirror states obtained by inverting the momentum flows.

\section{Conclusions}
A modified Grover’s quantum algorithm, for the identification of causal singular configuration of multiloop Feynman integrals was described.
The algorithm was illustrated with the six-edge two-eloop topology, implemented through the quantum simulator provided by IBM Qiskit, and successfully determined all the causal singular configurations.

The development of this algorithm represents a relevant resource to the LTD framework, providing an efficient procedure to search the causal singular configurations required to bootstrap the causal LTD representation of mutiloop scattering amplitudes.

\section*{Acknowledgments}
I want to thank G. Rodrigo for the guidance through the development of this work, and A. Renter\'ia for the companionship and support. 
Support for this work has been received in part by MCIN/AEI/10.13039/501100011033, Grant No. PID2020-114473GB-I00, Consejo Nacional de Ciencia y Tecnología and Universidad Aut\'onoma de Sinaloa.

\end{document}